\documentclass[aps,prc,twocolumn,superscriptaddress,showpacs]{revtex4-2}

\usepackage[pdftex]{graphicx}
\usepackage{amsmath} 
\usepackage{amssymb} 
\usepackage{mathtools}
\usepackage{color} 
\usepackage{textcomp} 
\usepackage{lmodern} 
\usepackage[T1]{fontenc}

\begin{document}

\title{Incoherent approximation for neutron up-scattering cross sections and its corrections for slow neutrons and low crystal temperatures}

\author{Stefan D\"{o}ge}
\email[Corresponding author Stefan Doege: ]{stefan.doege@tum.de}
\affiliation{Technische Universit\"{a}t M\"{u}nchen, Department of Physics E18, James-Franck-Strasse 1, D-85748 Garching, Germany}
\affiliation{Institut Laue--Langevin, 71 avenue des Martyrs, F-38042 Grenoble Cedex 9, France}

\author{Chen-Yu Liu}
\affiliation{Department of Physics, Indiana University, 727 East Third Street, Bloomington, Indiana 47405, USA}

\author{Albert Young}
\affiliation{Department of Physics, North Carolina State University, Raleigh, North Carolina 27695, USA}

\author{Christoph Morkel}
\affiliation{Technische Universit\"{a}t M\"{u}nchen, Department of Physics E21, James-Franck-Strasse 1, D-85748 Garching, Germany}

\date{\today}

\begin{abstract}
The incoherent approximation (IA) is often used for calculating the one-phonon inelastic neutron scattering cross section for arbitrary solids. It is valid for thermal neutrons but for slow neutrons it requires a correction, which is significant for isotopes that are strong coherent scatterers. In this article, we present the extension of the Placzek--Van Hove corrections for slow neutrons in the limit of low temperatures using the example of solid \emph{ortho}-deuterium (sD$_2$). Our approach yields realistic one-phonon up-scattering cross sections for sD$_2$ and shows the IA to be a factor of 2 to 5 too high for ultracold neutron (UCN) up-scattering in sD$_2$. Our calculations are compared with previously published Monte Carlo calculations of the one-phonon cross section based on the dynamic structure function $S(q,\omega)$ of polycrystalline \emph{ortho}-deuterium and are found to be consistent with them. Furthermore, we provide the means for easily replicable calculations of the one-phonon up-scattering cross sections of solid \emph{ortho}-deuterium for slow neutrons. These should from now on be used in calculations and simulations of UCN scattering in sD$_2$.
\\
\\
Published online on 10 May 2021: \url{https://doi.org/10.1103/PhysRevC.103.054606} \\ S. D\"{o}ge, C.-Y. Liu, A. Young, C. Morkel, Physical Review C 103 (5), 054606 (2021) \\ \textcopyright\, 2021. This manuscript version is made available under the \href{https://creativecommons.org/licenses/by/4.0/}{CC-BY 4.0 license}.
\end{abstract}

\maketitle

\section{Introduction}

The slowest neutrons -- ultracold neutrons (UCNs) -- can be produced by various methods. One of them is the \emph{superthermal} production process, in which a cold neutron excites a phonon or rotational transition in a solid or liquid converter medium and becomes so slow that it falls in the velocity range of UCNs, from 0 to about 10~m/s. The UCNs produced need to be extracted from the converter within their lifetime to achieve a maximum UCN flux available for experiments, most of which are in the fields of particle and astrophysics. According to Golub and Pendlebury~\cite{golub:1975}, this lifetime of UCNs inside the converter is determined by (i) the absorption of the neutron by a nucleus of the converter medium, (ii) loss of the neutron on the converter wall, and (iii) inelastic up-scattering.

The inelastic up-scattering cross section of \emph{ortho}-deuterium for UCNs is thus an important parameter in the calculations of UCN extraction rates from converters based on solid \emph{ortho}-deuterium (sD$_2$). In sD$_2$, the vibrational levels are not populated and of the rotational levels $J$ only the lowest, the \emph{ortho} level ($J = 0$), is populated in thermal equilibrium at temperatures below 18~K~\cite{silvera:1980}. Therefore, a UCN can absorb energy -- and hence be \emph{up-scattered} to a higher energy range -- only from a phonon in the material. For the up-scattering of UCNs, only one-phonon inelastic scattering plays a role. According to calculations by Placzek~\cite{placzek:1954,turchin:1965}, two- and multi-phonon processes can be neglected.

The incoherent approximation (IA) is often used for calculating the inelastic neutron scattering cross section for arbitrary solids by equating coherent with incoherent scattering, and neglecting the interference term in the correlation function of the double-differential scattering cross section. However, for slow neutrons, it requires a correction for the neglected interference term, especially for isotopes with a strong coherent contribution to the scattering process. In the following, we present such corrections for solid \emph{ortho}-deuterium in the UCN limit and show that the uncorrected IA, which has hitherto been used to explain the entire UCN scattering cross section~\cite{atchison:2005-sol}, is a factor of 2 to 5 too high. This finding sheds new light on the existence of crystal defects in sD$_2$~\cite{doege:2019-phd}, which must be taken into account in the design and upgrade of UCN converters based on sD$_2$.

\section{Deuterium Molecular Cross Sections}

Hamermesh and Schwinger~\cite{hamermesh:1946} developed models to calculate the scattering cross sections of \emph{free} deuterons and \emph{interaction-free} deuterium molecules for various spins (\emph{ortho} and \emph{para} species), vibrational, and rotational transitions. These cross sections are \emph{self} cross sections, meaning they do not take into account collective dynamics of the deuterium sample but only the interaction of the neutron with a single scattering particle. They calculated these cross sections for thermal and subthermal neutrons in deuterium gas at low temperatures.

Later, Young and Koppel~\cite{young-koppel:1964} calculated the double-differential scattering cross sections of \emph{ortho}- and \emph{para}-deuterium for a wider temperature and energy range, taking into account rotation, vibration, and translation of the molecule. The latter was treated as unperturbed by neighboring molecules. The special case of a liquid was also described but is applicable only to neutron energies above the Debye temperature of the sample. For \emph{ortho}-deuterium, the Debye temperature is around $\Theta_\text{D}=110$~K, i.e., 9.48~meV, at low sample temperatures of $T=0\ldots18$~K~\cite{hill:1959,nielsen:1973}.

\section{The Incoherent Approximation for Thermal Neutrons}\label{sect:incoh-approx}

One-phonon inelastic scattering can be both coherent and incoherent. Both can be separately described and contribute to the total double-differential cross section~\cite{turchin:1965, ignatovich:1990, golub:1991}, $\text{d}^2 \sigma/\text{d}\Omega \text{d}E$,

\begin{multline}\label{eq:scattering-full}
\frac{\text{d}^2 \sigma^{\pm\text{1ph}}}{\text{d}\Omega \text{d}E} = \frac{1}{4\pi}\frac{k_\text{f}}{k_\text{i}} \left[\sigma^\text{coh} j_n^2\left(\frac{aq}{2}\right) S_\text{coh}^\text{1ph}(q,\omega) \right.\\
\left. + \sigma^\text{inc} j_n^2\left(\frac{aq}{2}\right) S_\text{inc}^\text{1ph}(q,\omega) \right].
\end{multline}

\noindent
The quantities $\sigma^\text{coh}$ and $\sigma^\text{inc}$ represent the coherent and incoherent molecular scattering cross section of a given rotational transition of the deuterium molecule ($J=0\rightarrow J'=0, 1\,{\rightarrow}\, 1, 1\,{\rightarrow}\, 0$). The spherical Bessel functions $j_n^2 \left(\frac{aq}{2}\right)$ represent the shape of the deuterium molecule in its different rotational states~\cite{young-koppel:1964}, $k_\text{i}$ is the wave vector of the incoming neutron, and $k_\text{f}$ that of the scattered neutron. The momentum transfer to the scattering neutron is $q = k_\text{i}-k_\text{f}$, and $\hbar\omega=E_\text{i}-E_\text{f}$ is the sample's energy change. Variable $a$ is the equilibrium distance of the two deuterons in the deuterium molecule. The collective dynamics of the scattering system are included in the coherent and incoherent scattering functions $S_\text{coh}(q,\omega)$ and $S_\text{inc}(q,\omega)$, respectively. Both are related to integrals of correlation functions, which are explained in textbooks~\cite{turchin:1965, ignatovich:1990, golub:1991} and shall not be repeated here.

When the coherent and incoherent double-differential cross sections are added up to calculate the total, the first term is a correlation function with a structure the same as that for incoherent scattering. Approximating coherent scattering in Eq.~\ref{eq:scattering-full} by the incoherent scattering law -- in other words, all scattering events are treated as uncorrelated and interference effects are neglected -- leads to a much simpler expression. The incoherent scattering law is then weighted with the sum of the coherent and incoherent molecular scattering cross sections, $\sigma^\text{sc}=\sigma^\text{coh}+\sigma^\text{inc}$. This simplifying approach is called the \emph{incoherent approximation} (IA)~\cite{turchin:1965},

\begin{equation}\label{eq:incoh-approx-basic}
\frac{\text{d}^2 \sigma^{\pm\text{1ph}}}{\text{d}\Omega \text{d}E} \simeq \frac{1}{4\pi}\frac{k_\text{f}}{k_\text{i}} \left[ \sigma^\text{sc} j_n^2\left(\frac{aq}{2}\right) S_\text{inc}^\text{1ph}(q,\omega) \right].
\end{equation}

\noindent
It is only applicable when the wavelength of the incoming neutron is not larger than the intermolecular distances in the scattering system~\cite{placzek:1951}. These distances of a few angstroms correspond to neutron energies of a few meV (subthermal neutrons). In other words, for scattering of very cold and ultracold neutrons, the IA is not valid and interference effects need to be taken into account -- for homogeneous substances as well as for substances with long-range order~\cite{turchin:1965}.

Inserting into Eq.~\ref{eq:incoh-approx-basic} the textbook result for $S_\text{inc}^\text{1ph}(q,\omega)$ for cubic crystals, e.g., from Gurevich and Tarasov~\cite{tarasov-gurevich:1968} or Lovesey~\cite{lovesey:1986}, yields
\begin{equation}\label{eq:incoh-approx-gurevich}
\frac{\text{d}^2 \sigma^{\pm\text{1ph}}}{\text{d}\Omega \text{d}E} = \frac{\sigma^\text{sc}}{4\pi}\frac{k_\text{f}}{k_\text{i}} e^{-2W(q)} \frac{q^2}{2M} \frac{Z(\omega)}{\omega} 
\begin{cases}
  n(\omega) + 1, & \omega > 0\\
  n(\omega), & \omega < 0
\end{cases}
\end{equation}

\noindent
Here, $\text{exp}[-2W(q)]$ is the Debye--Waller factor with $2W(q) = \left\langle u^2(t)\right\rangle q^2/3$ and $\left\langle u^2(t)\right\rangle$ including the zero-point motion of the scattering molecules of mass $M$ at low temperatures~\cite{squires:1978}. The energy transfer direction is given by $\hbar\omega$. For energy gain of the scattered slow neutron, i.e., energy loss of the sample ($\hbar\omega <0$), the Bose factor is $n(\omega) = [\text{exp}(\hbar\omega/k_\text{B}T)-1]^{-1}$ with $k_\text{B}$ the Boltzmann constant. This neutron up-scattering by phonon annihilation is known to be the main loss channel for UCNs in sD$_2$ at low temperatures.

Using the Debye approximation for the density of states (DOS), $Z(\omega) = 3\omega^2 / \omega_\text{D}^3$ for $\omega \leq \omega_\text{D}$, with $\omega_\text{D}$ the Debye frequency of the solid, the integration of the phonon annihilation part of Eq.~\ref{eq:incoh-approx-gurevich} (superscript \emph{-1ph}) over the kinematic region yields~\cite{tarasov-gurevich:1968}
\begin{multline}\label{eq:incoh-approx-tarasov-gurevich}
\sigma_{00}^\text{-1ph} = 24 \sigma_{00}^\text{sc} \left( \frac{m_\text{n}}{M} \right) \left(\frac{E_\text{i}}{k_\text{B}\Theta_\text{D}} \right)^3 \\
\times \int \int \text{d}\xi \text{d}\eta e^{-2W(\eta)} \frac{\eta^3 \xi j_0^2(ak_\text{i} \eta)}{\exp{(\xi E_\text{i}/k_\text{B}T)}-1},
\end{multline}

\noindent
with the dimensionless variables $\eta \equiv \frac{q}{2k_\text{i}}$ and $\xi \equiv \frac{\hbar \omega}{E_\text{i}}$. $E_\text{i}$ represents the kinetic energy of the neutron before the scattering event and $m_\text{n}$ stands for the neutron mass. The spherical Bessel function $j_0^2 (ak_\text{i} \eta)$ was added to the equation to account for the shape of the \emph{ortho}-deuterium molecule in its rotational ground state. Phonon creation can be neglected in the case of UCNs.

Since the kinematic region is narrow (see Fig.~\ref{fig:ucn-parabola-coh-phonons}), the Debye--Waller factor (DWF) and the Bessel function can be calculated for an average value of their integration variables. In doing so, they become constants in $\eta$ and can be pulled out of the integral over $\text{d}\eta$. Now the inner integral over $\eta$ can be solved analytically, which yields~\cite{doege:2019-phd}
\begin{flalign}
\sigma_{00}^\text{-1ph} = &24 \sigma_{00}^\text{sc} \left( \frac{m_\text{n}}{M} \right) \left(\frac{E_\text{i}}{k_\text{B}\Theta_\text{D}}\right)^3 \nonumber \\
&\times \int_0^{k_\text{B}\Theta_\text{D}/E_\text{i}} \text{d}\xi e^{-2\overline{W}(\overline{\eta})} \frac{1}{8} \sqrt{1+\xi} (2+\xi) \nonumber \\
&\times \left[ j_0^2\left( \frac{ak_\text{i}\sqrt{1+\xi}}{2} \right) \frac{\xi}{\exp{(\xi E_\text{i}/ k_\text{B}T)}-1} \right].
\end{flalign}

\noindent
This integral has to be solved numerically and yields the full incoherent approximation.

For the limit of $T \rightarrow 0$ and $E_\text{i} \rightarrow 0$ (very slow neutrons), an analytical solution is known as the Stepanov equation~\cite{stepanov:1975},
\begin{equation}
\sigma^\text{-1ph} \simeq \frac{45}{8}\zeta \left(\frac{7}{2}\right) \sqrt{\pi}\sigma_{00}^\text{sc}\left(\frac{m_\text{n}}{M}\right) \sqrt{\frac{k_\text{B}T}{E_0}} \left(\frac{T}{\Theta_\text{D}}\right)^3,
\end{equation}

\noindent
where $\zeta (x)$ is the Riemann zeta function~\cite{abramowitz-stegun:1965}.

This equation neglects the exact forms of the Debye--Waller factor (DWF) and the Bessel function $j_0^2$ by setting both of them equal to 1. Furthermore, the upper integration limit is taken as infinity instead of the actual energy relation, which greatly simplifies the integration of the double-differential cross section over the kinematic region; it can now be carried out analytically. This simplification, however, comes at the price of limited validity. The Stepanov equation is only valid at low temperatures ($T<5$~K for deuterium) and low neutron energies. At a deuterium temperature of $T=18$~K, the Stepanov equation delivers one-phonon cross sections about 40\% above the full IA (see Table~\ref{tab:inc-approx-comparison}). Therefore, it is not suited for calculating the one-phonon up-scattering cross section in ``warm'' deuterium crystals of $T=10\ldots 18$~K.

\section{The Placzek--Van Hove Correction}

Placzek and Van Hove (PVH) developed a detailed approach~\cite{placzek:1955} to account for the interference term arising from coherent scattering, which is neglected by the incoherent approximation. In this section, we apply the ``PVH correction'' to scattering cross sections calculated using the IA for thermal and slow neutron scattering in samples at room temperature, and compare our results with calculations by others as well as experimental results -- to verify our correct understanding of it. Eventually, we expand the PVH correction to the limit of low temperatures and calculate correction terms for sD$_2$. Then we compare the corrected one-phonon scattering cross sections of sD$_2$ for UCNs with other calculation methods.

In the case of \emph{thermal} neutrons, the IA is valid for large momentum transfers $q$, as shown by Oskotskii~\cite{oskotskii:1967} and Schober~\cite{schober:2014}. Corrections to it are of the order of $(\lambda/d)^2$, with $d$ the interatomic distances in the lattice of the scattering system and $\lambda$ the neutron wavelength~\cite{placzek:1951,turchin:1965}. Calculating these corrections for copper and aluminum, we found only marginal correction terms of $-1.6$\% (Cu) and $-0.1$\% (Al). Hence, the IA is fully justified for thermal neutrons.

In the \emph{slow} neutron limit, we calculated PVH corrections for various metals at room temperature. For copper, we found a correction to the IA of $-8$\%, and for aluminum of $+11$\%~\cite{doege:2020-foils}. These values are in line with those published earlier for slow neutrons~\cite{placzek:1955,turchin:1965}.

Next, we used the general approach of PVH to consider the limit of low temperatures and low neutron energies in the range below 1~$\mu$eV ($v<15$~m/s), and applied the results to solid \emph{ortho}-deuterium. We only repeat equations from PVH's paper where necessary for understanding.

\subsection{General formulas}

Using the notation of PVH, Eq.~\ref{eq:scattering-full} can be written as
\begin{equation}
\sigma^\text{-1ph} = \sigma^\text{inc} S^\text{inc} + \sigma^\text{coh} S^\text{coh},
\end{equation}
\noindent
where $S^\text{coh}$ and $S^\text{inc}$ stand for the coherent and incoherent scattering law \emph{integrated} over the kinematic region. Introducing $\delta S = S^\text{coh}-S^\text{inc}$ and expanding this equation to $\sigma^\text{-1ph} = (\sigma^\text{coh}+\sigma^\text{inc}) S^\text{inc} + \sigma^\text{coh} \delta S$, we obtain
\begin{equation}\label{eq:placzek-correction}
\sigma^\text{-1ph} = \underbrace{\sigma^\text{sc} S^\text{inc}}_{\sigma^\text{IA}} \underbrace{\left[ 1+\frac{\sigma^\text{coh}}{\sigma^\text{sc}} \left( \frac{\delta S}{S^\text{inc}}\right) \right]}_{\text{correction factor }\chi},
\end{equation}

\noindent
where $\sigma^\text{IA}$ represents the IA and the second factor is the correction factor $\chi$ arising from interference effects due to the collective dynamics in the solid. For \emph{ortho}-deuterium $\sigma^\text{coh}=\sigma_\text{00}^\text{coh}$ and $\sigma^\text{sc}=\sigma_\text{00}^\text{sc}$. $\delta S/S^\text{inc}$ is the Placzek--Van Hove correction term and the key parameter to be calculated.

\subsection{Qualitative discussion}

\begin{figure}[!ht]
\centering
\includegraphics[width=1.00\columnwidth]{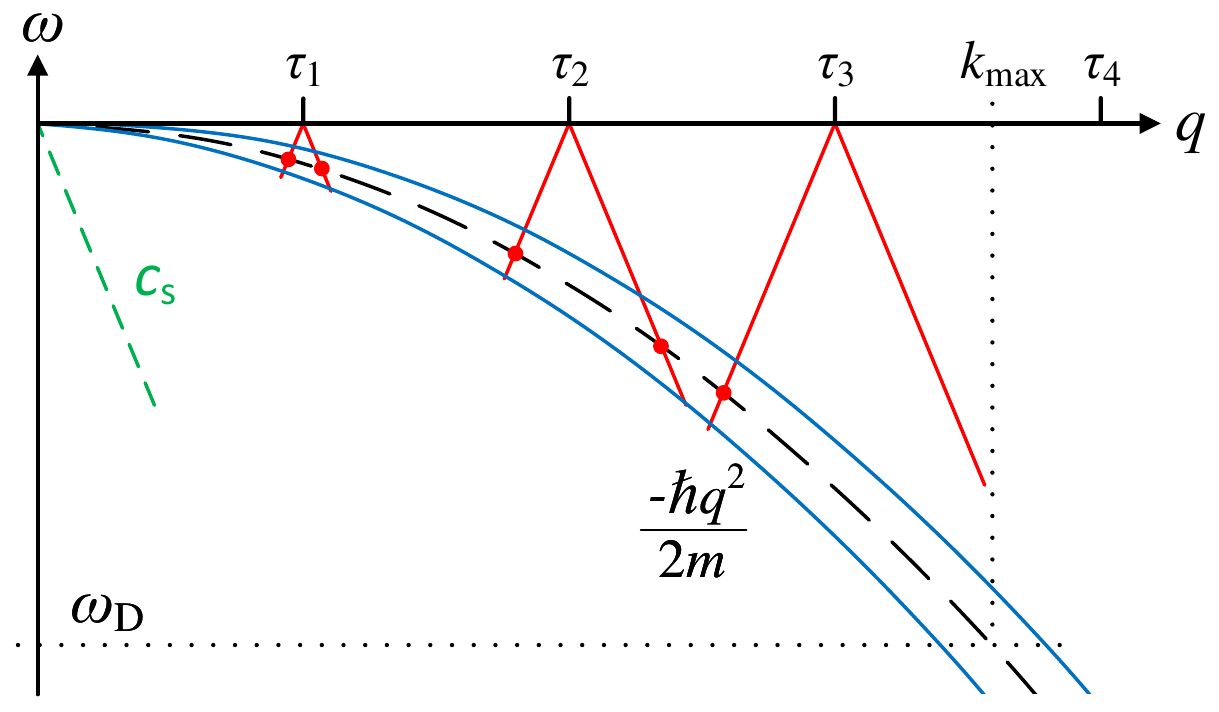}
\caption[UCN parabola with lattice vectors]{\label{fig:ucn-parabola-coh-phonons}Kinematic region and coherent UCN up-scattering in the Debye approximation after Turchin~\cite{turchin:1965}. Lattice vectors $\tau_i$ beyond $k_\text{max} \approx (\frac{2 m \omega_\text{D}}{\hbar})^{1/2}$, with $\omega_\text{D}$ the Debye frequency, do not contribute to the coherent scattering. $\omega = c_\text{s} q$ with $c_\text{s}$ the sound velocity. The dispersion curve of the free neutron is given by $\omega = -\hbar q^2/2m_\text{n}$.}
\end{figure}

Before we evaluate the correction term quantitatively, it seems instructive to qualitatively discuss the contributions to $\sigma^\text{-1ph}$. To this end, Fig.~\ref{fig:ucn-parabola-coh-phonons} shows the relevant part of the kinematic region for one-phonon up-scattering of UCNs ($\omega < 0$).

All \emph{incoherent} one-phonon up-scattering events fall between the two limiting blue parabolas representing forward and backward scattering of UCNs in the $q$--$\omega $ plane. Both parabolas follow closely the free-particle dispersion curve of the neutron (black dashed curve). In the Debye approximation (DA), as used here, the ordinate is limited at $\lvert\omega\rvert \leq \omega_\text{D}$, with $\omega_\text{D}=k_\text{B}\Theta_\text{D}/\hbar$. As a consequence, the relevant range on the abscissa is also limited, at $k_\text{max} \approx (2m_\text{n}\omega_\text{D}/\hbar)^{1/2}$ (the dotted straight lines in Fig.~\ref{fig:ucn-parabola-coh-phonons}). Beyond these points, no one-phonon up-scattering is possible and thus the integration of Eq.~\ref{eq:incoh-approx-tarasov-gurevich} needs to be carried out over the kinematic region between the parabolas up to $\omega_\text{D}$ and $k_\text{max}$.

The \emph{coherent} one-phonon scattering events emanate from the intersections of the sound dispersion $\omega = c_\text{s} q$ ($c_\text{s}$ is the sound velocity of the medium in DA, red lines in Fig.~\ref{fig:ucn-parabola-coh-phonons}) -- originating at various lattice vectors $\tau_i$ -- with the free UCN dispersion curve $\omega = -\hbar q^2 / 2m_\text{n}$. Two features of Fig.~\ref{fig:ucn-parabola-coh-phonons} are worth noting: (i) Only a limited number of lattice vectors $\tau_i$ are ``active'' as Bragg peaks and fulfill momentum and energy conservation simultaneously for UCNs~\cite{placzek:1955,golub:1991}, and (ii) the crossings of the phonon dispersion with the kinematic region (red dots in Fig.~\ref{fig:ucn-parabola-coh-phonons}) stand, in the case of UCNs, for a very small area of width $\Delta q \propto k_\text{i}$ and $\Delta \omega \propto E_\text{i}$. Hence, together with the limited number of Bragg peaks, the integration area in the coherent case is only a small fraction of the whole kinematic region. This means that, in the case of UCNs, coherent phonon scattering makes a much smaller contribution to the total one-phonon up-scattering than incoherent events. For UCNs, the obvious conclusion from this is that the IA significantly overestimates the total one-phonon up-scattering cross section.

\subsection{Outline of the Placzek--Van Hove theory and expansion to low temperatures}

In the following quantitative calculation we adopt reduced variables for wave vectors $\vec{q}'$, lattice vectors $\vec{\tau}'$, and temperature $T'$ as used by PVH,
\begin{equation}
\vec{q}' = \frac{\vec{q}}{q_\text{D}}, \vec{\tau}' = \frac{\vec{\tau}}{q_\text{D}}, T' = \frac{T}{\Theta_\text{D}},
\end{equation}
\noindent
with the Debye vector $q_\text{D} = (6\pi^2n_0)^{1/3}$, $n_0$ the number density of molecules in the crystal, and $\Theta_\text{D}$ its Debye temperature.

Using these variables, the scattering surface becomes (see Eq.~5.12 in PVH~\cite{placzek:1955})
\begin{equation}
(\vec{\tau}' + \vec{q}')^2 = f q'
\end{equation}
with
\begin{equation}
f = \left( \frac{k_\text{max}}{q_\text{D}} \right)^2
\end{equation}
resulting in
\begin{equation}
\tau_\text{max}' \lesssim \sqrt{f}+1.\label{eq:tau-max}
\end{equation}
$f$ is the crucial parameter of the PVH theory and shall be called the fill factor because $f^{3/2}$ is defined as the ratio of the energetically accessible $k$ space to the volume of a Brillouin zone. For a large variety of elements $f \lesssim 10$, which means that only a few lattice vectors of finite length $\tau'_i$ are involved in coherent UCN scattering processes.

The correction term $\delta S /S^\text{inc}$ for $k_\text{i}\rightarrow 0$ (UCN limit) can be expressed as
\begin{equation}\label{eq:pvh-term}
\frac{\delta S}{S^\text{inc}} = -1 + \frac{1}{N} \sum^{\tau'_\text{max}}_{\tau'_i} \frac{1}{2f^{1/2}\tau'} \int^{q'_\text{max}}_{q'_\text{min}} g(\tau') \Psi (q',T') q' \text{d}q',
\end{equation}
\noindent
according to PVH~\cite{placzek:1955}, Eq.~5.20. Here, the positive term represents the coherent contribution to $\delta S /S^\text{inc}$, $g(\tau')$ is the number of vectors of length $\tau'$, i.e., the multiplicity of $\tau'$~\cite{placzek:1951}, while $\Psi (q',T')$ describes the product of the DWF and the Bose factor $n(\omega)$ in PVH notation,
\begin{equation}
\Psi (q',T') = \frac{e^{-6m/M\times\phi (T')q'}}{e^{q'/T'}-1}.
\end{equation}
\noindent
$\phi (T')$ is the effective temperature of the scattering system. PVH considered only the high-temperature limit of $\phi (T')$. We calculated it for the quantum crystal sD$_2$ at low $T'$, including quantum effects, such as zero-point motion at low temperatures. The reduced vectors $q'_\text{min,max}$ are given by PVH in their Eqs.~5.18 and 5.19, respectively. The normalization factor $N$ in Eq.~\ref{eq:pvh-term} stands for the incoherent approximation in PVH notation,
\begin{equation}
N = 3\int^1_0 \Psi (q', T')q'^{5/2} \text{d}q'.
\end{equation}

In summary, the equations of the PVH correction given in this section allow for the calculation of the interference correction term $\delta S/ S^\text{inc}$, now also for low temperatures.

It should be mentioned that Binder~\cite{binder:1970} worked out a calculation of $\delta S$ ($k_\text{i}\rightarrow 0$) in real-space representation. However, his ansatz shows an extremely poor convergence in comparison to the Fourier-space representation of PVH~\cite{placzek:1955}.

\subsection{Parameters of sD$_2$ for quantitative calculations}

In this section, we give the parameters used in our calculations of the corrections to the IA for solid \emph{ortho}-deuterium. All parameters, other than the reduced temperature $T'$, were regarded to be independent of temperature as they change only little from 5 to 18~K, which is the relevant temperature range for sD$_2$. The parameters are as follows: mass ratio $M_\text{D2}/m_\text{n} = 4$, number density $n_0 = 3.0\times 10^{22}$~cm$^{-3}$, Debye temperature $\Theta_\text{D}=70$~K (only for acoustic phonon modes, see explanation below), Debye vector $q_\text{D}=1.21$~\AA$^{-1}$, $k_\text{max}=1.71$~\AA$^{-1}$, and the fill factor $f = 1.98$.

These parameters are quasi-independent of the solid's structure, whether it be an fcc or bcc crystal~\cite{placzek:1955}. The PVH approach is only valid for Bravais lattices, i.e., with one atom per primitive cell. Real deuterium, however, has an hcp lattice with $a = 3.60$~\AA{} and an ideal $c/a$ ratio of 1.63~\cite{mucker:1968}, giving a volume of $V_\text{p.c.}=65.86$~\AA$^3$ for the primitive cell with two deuterons, and a number density of $3.04\times 10^{22}$~cm$^{-3}$. An fcc structure of the same number density has a lattice constant of $a_\text{fcc}=5.09$~\AA. In fact, deuterium assumes an fcc structure with $a = 5.084\pm 0.004$~\AA{} below 2.8~K~\cite{mucker:1966}. Hence, the treatment of real sD$_2$ as its equivalent fcc structure is justified.

In the fcc structure, only the smallest two reduced lattice vectors of lengths $\tau'_\text{1/2}$ fulfill Eq.~\ref{eq:tau-max}, and are thus the only ones to be taken into account with their respective multiplicities.

As Liu \textit{et al.}~\cite{liu:2010} have shown, the DOS of sD$_2$ has two peaks, one at $\Theta_\text{D}^\text{ac}=70$~K due to acoustic phonons and one at $\Theta_\text{D}^\text{opt}=110$~K due to optical phonons. For the PVH treatment of sD$_2$, we used the reduced Debye temperature of 70~K, since the approach is only strictly applicable to Bravais-type crystals, i.e., lattices without optical phonon modes. 

To compensate the neglect of the real hcp structure of sD$_2$, which -- in contrast to fcc -- contains optical phonon modes, we consider the simple relation $\sigma^\text{-1ph}_\text{coh}\propto \sqrt{\omega} n(\omega)$ for the coherent UCN scattering cross section of phonons~\cite{golub:1991}. For sD$_2$, the contribution of optical phonons to the scattering is related to the contribution of acoustic phonons as follows
\begin{equation}
\frac{\sigma^\text{opt}}{\sigma^\text{ac}} = \sqrt{\frac{\omega_\text{D}^\text{opt}}{\omega_\text{D}^\text{ac}}} \frac{n(\omega_\text{D}^\text{opt})}{n(\omega_\text{D}^\text{ac})} = 1.254 e^{-\Delta\Theta_\text{D}/T},
\end{equation}
\noindent
where $\Delta\Theta_\text{D}$ is the difference between the Debye temperatures of the optical and acoustic phonons.

From this estimate, it can be seen that the optical phonons contribute little to the coherent scattering at low temperatures due to their small occupancy factor. For sD$_2$ at 5~K, the optical phonons cause less than one permille additional scattering, at 18~K they contribute an extra 14\%. We multiplied the correction factor $(1+\sigma^\text{opt}/\sigma^\text{ac})$ to the second term of Eq.~\ref{eq:pvh-term}, thus taking optical phonons into account for the $\delta S/S^\text{inc}$, $\sigma^\text{-1ph}$, and $\chi_\text{D2}$ values in Table~\ref{tab:inc-approx-comparison}.

\section{Comparison of Results Using Different Calculation Methods}

Placzek and Van Hove provided a way to calculate $\frac{\delta S}{S^\text{inc}}$, which we applied and expanded to the limit of low temperatures. The results of our calculations of $\frac{\delta S}{S^\text{inc}}$ for sD$_2$ using the parameters listed above, as well as various other means of calculating the one-phonon up-scattering, are shown and compared with one another in Table~\ref{tab:inc-approx-comparison} for a selection of typical solid deuterium temperatures and a UCN (in-medium) velocity of 10~m/s.

It is obvious that the IA used hitherto yields cross sections $\sigma^\text{IA}$ that are, depending on the temperature, a factor of 2 to~5 too high. Such values for sD$_2$ were used, for example, by Atchison \textit{et al.}~\cite{atchison:2005-ucnprod,atchison:2005-sol}.

Our \emph{corrected} one-phonon up-scattering cross sections $\sigma^\text{-1ph}$ lie less than 14\% below those of Liu \textit{et al.}~\cite{liu:2010}, who used a first-principles Monte Carlo calculation of the dynamic structure function $S(q,\omega)$ of polycrystalline sD$_2$ to obtain their results. The excellent agreement of the results from these two very different approaches is a mutual validation and underlines the pertinence of the corrections to the IA presented here. The results of both approaches are furthermore supported by scattering cross section data obtained from experiments in Ref.~\cite{doege:2019-phd}.

\begin{table}[!ht]
\centering
  \begin{tabular}{c c c c c c c}\hline\hline
Temp. & $\frac{\delta S}{S^\text{inc}}$ & $\sigma^\text{Stepanov}$ & $\sigma^\text{IA}$ & $\sigma^\text{-1ph}$ & $\chi_\text{D2} = $ & $\chi^\text{Liu}_\text{D2}$ \\
{[K]} & & [barn] & [barn] & [barn] & $\sigma^\text{-1ph}/\sigma^\text{IA}$ & \\\hline
5 & -0.9915 & 0.21 & 0.19 & 0.04 & 0.19 & 0.22 \\
10 & -0.8381 & 2.36 & 1.99 & 0.63 & 0.32 & 0.37 \\
15 & -0.6679 & 9.74 & 7.39 & 3.40 & 0.46 & 0.50 \\
18 & -0.5904 & 18.4 & 12.9 & 6.70 & 0.52 & 0.54 \\\hline\hline
  \end{tabular}
\caption[Placzek--Van Hove correction term]{The Placzek--Van Hove correction terms $\frac{\delta S}{S^\text{inc}}$ calculated for solid \emph{ortho}-deuterium are given for various temperatures. Inserting these numbers along with the nuclear cross sections into Eq.~\ref{eq:placzek-correction} allows us to calculate $\sigma^\text{-1ph}$ -- the corrected incoherent approximation for the scattering cross section of sD$_2$. The ratios of corrected IA $\sigma^\text{-1ph}$ to ``old'' IA $\sigma^\text{IA}$ are given as $\chi_\text{D2}$. These ratios compare well with those obtained by Liu \textit{et al.}~\cite{liu:2010}, $\chi^\text{Liu}_\text{D2}$. The scattering cross sections $\sigma^\text{Stepanov}$ yielded by the Stepanov formula are shown for comparison. All cross sections are given for one \emph{molecule} and have been calculated for a UCN velocity of 10~m/s. They scale with 1/$v$ for other velocities $v$ in the UCN range, i.e., $\sigma v = \text{const.}$.}\label{tab:inc-approx-comparison}
\end{table}

\section{Conclusion}

The widely used incoherent approximation for neutron one-phonon scattering cross sections was shown to be a factor of 2 to~5 too high for ultracold neutron (UCN) up-scattering in solid \emph{ortho}-deuterium (sD$_2$). Applying a correction for interference effects based on the approach of Placzek and Van Hove~\cite{placzek:1955}, which we expanded to low temperatures, yielded much lower, realistic one-phonon scattering cross sections. These should from now on be used in calculations and simulations of UCN up-scattering in sD$_2$. Our results agree very well with Monte Carlo simulations by Liu \textit{et al.}~\cite{liu:2010}. This agreement is a mutual confirmation for both results as they were obtained using completely different methods. Both are furthermore supported by experimental results in Ref.~\cite{doege:2019-phd}.

\begin{acknowledgments}
We thank Brian C. Dye from the University of Hawaii at Manoa for a critical reading of the manuscript. This work received funding from FRM~II/ Heinz Maier-Leibnitz Zentrum (MLZ), Munich, Germany. The open access publication fee was paid for by the University Library and the Physics Department of the Technische Universit\"at M\"unchen, Munich, Germany.

The results for this paper were produced as part of the Ph.D. thesis of Stefan D\"oge~\cite{doege:2019-phd}.
\end{acknowledgments}

\bibliography{Bibliography}

\begin{thebibliography}{27}%
\makeatletter
\providecommand \@ifxundefined [1]{%
 \@ifx{#1\undefined}
}%
\providecommand \@ifnum [1]{%
 \ifnum #1\expandafter \@firstoftwo
 \else \expandafter \@secondoftwo
 \fi
}%
\providecommand \@ifx [1]{%
 \ifx #1\expandafter \@firstoftwo
 \else \expandafter \@secondoftwo
 \fi
}%
\providecommand \natexlab [1]{#1}%
\providecommand \enquote  [1]{``#1''}%
\providecommand \bibnamefont  [1]{#1}%
\providecommand \bibfnamefont [1]{#1}%
\providecommand \citenamefont [1]{#1}%
\providecommand \href@noop [0]{\@secondoftwo}%
\providecommand \href [0]{\begingroup \@sanitize@url \@href}%
\providecommand \@href[1]{\@@startlink{#1}\@@href}%
\providecommand \@@href[1]{\endgroup#1\@@endlink}%
\providecommand \@sanitize@url [0]{\catcode `\\12\catcode `\$12\catcode
  `\&12\catcode `\#12\catcode `\^12\catcode `\_12\catcode `\%12\relax}%
\providecommand \@@startlink[1]{}%
\providecommand \@@endlink[0]{}%
\providecommand \url  [0]{\begingroup\@sanitize@url \@url }%
\providecommand \@url [1]{\endgroup\@href {#1}{\urlprefix }}%
\providecommand \urlprefix  [0]{URL }%
\providecommand \Eprint [0]{\href }%
\providecommand \doibase [0]{https://doi.org/}%
\providecommand \selectlanguage [0]{\@gobble}%
\providecommand \bibinfo  [0]{\@secondoftwo}%
\providecommand \bibfield  [0]{\@secondoftwo}%
\providecommand \translation [1]{[#1]}%
\providecommand \BibitemOpen [0]{}%
\providecommand \bibitemStop [0]{}%
\providecommand \bibitemNoStop [0]{.\EOS\space}%
\providecommand \EOS [0]{\spacefactor3000\relax}%
\providecommand \BibitemShut  [1]{\csname bibitem#1\endcsname}%
\let\auto@bib@innerbib\@empty
\bibitem [{\citenamefont {Golub}\ and\ \citenamefont
  {Pendlebury}(1975)}]{golub:1975}%
  \BibitemOpen
  \bibfield  {author} {\bibinfo {author} {\bibfnamefont {R.}~\bibnamefont
  {Golub}}\ and\ \bibinfo {author} {\bibfnamefont {J.~M.}\ \bibnamefont
  {Pendlebury}},\ }\bibfield  {title} {\bibinfo {title} {Super-thermal sources
  of ultra-cold neutrons},\ }\href
  {https://doi.org/https://doi.org/10.1016/0375-9601(75)90500-9} {\bibfield
  {journal} {\bibinfo  {journal} {Physics Letters A}\ }\textbf {\bibinfo
  {volume} {53}},\ \bibinfo {pages} {133} (\bibinfo {year} {1975})}\BibitemShut
  {NoStop}%
\bibitem [{\citenamefont {Silvera}(1980)}]{silvera:1980}%
  \BibitemOpen
  \bibfield  {author} {\bibinfo {author} {\bibfnamefont {I.~F.}\ \bibnamefont
  {Silvera}},\ }\bibfield  {title} {\bibinfo {title} {{T}he solid molecular
  hydrogens in the condensed phase: {F}undamentals and static properties},\
  }\href {https://doi.org/10.1103/RevModPhys.52.393} {\bibfield  {journal}
  {\bibinfo  {journal} {Rev. Mod. Phys.}\ }\textbf {\bibinfo {volume} {52}},\
  \bibinfo {pages} {393} (\bibinfo {year} {1980})}\BibitemShut {NoStop}%
\bibitem [{\citenamefont {Placzek}(1954)}]{placzek:1954}%
  \BibitemOpen
  \bibfield  {author} {\bibinfo {author} {\bibfnamefont {G.}~\bibnamefont
  {Placzek}},\ }\bibfield  {title} {\bibinfo {title} {Incoherent neutron
  scattering by polycrystals},\ }\href {https://doi.org/10.1103/PhysRev.93.895}
  {\bibfield  {journal} {\bibinfo  {journal} {Phys. Rev.}\ }\textbf {\bibinfo
  {volume} {93}},\ \bibinfo {pages} {895} (\bibinfo {year} {1954})}\BibitemShut
  {NoStop}%
\bibitem [{\citenamefont {Turchin}(1965)}]{turchin:1965}%
  \BibitemOpen
  \bibfield  {author} {\bibinfo {author} {\bibfnamefont {V.~F.}\ \bibnamefont
  {Turchin}},\ }\href@noop {} {\emph {\bibinfo {title} {Slow Neutrons}}},\
  {I}srael program for scientific translations\ (\bibinfo  {publisher} {Sivan
  Press, Jerusalem},\ \bibinfo {year} {1965})\ \bibinfo {note} {{R}ussian
  original: Medlennye nejtrony (Gosatomizdat, Moscow, 1963)}\BibitemShut
  {NoStop}%
\bibitem [{\citenamefont {Atchison}\ \emph
  {et~al.}(2005{\natexlab{a}})\citenamefont {Atchison}, \citenamefont {Blau},
  \citenamefont {van~den Brandt}, \citenamefont {Bry\ifmmode~\acute{s}\else
  \'{s}\fi{}}, \citenamefont {Daum}, \citenamefont {Fierlinger}, \citenamefont
  {Hautle}, \citenamefont {Henneck}, \citenamefont {Heule}, \citenamefont
  {Kirch}, \citenamefont {Kohlbrecher}, \citenamefont {K\"uhne}, \citenamefont
  {Konter}, \citenamefont {Pichlmaier}, \citenamefont {Wokaun}, \citenamefont
  {Bodek}, \citenamefont {Kasprzak}, \citenamefont {Ku\ifmmode~\acute{z}\else
  \'{z}\fi{}niak}, \citenamefont {Geltenbort},\ and\ \citenamefont
  {Zmeskal}}]{atchison:2005-sol}%
  \BibitemOpen
  \bibfield  {author} {\bibinfo {author} {\bibfnamefont {F.}~\bibnamefont
  {Atchison}}, \bibinfo {author} {\bibfnamefont {B.}~\bibnamefont {Blau}},
  \bibinfo {author} {\bibfnamefont {B.}~\bibnamefont {van~den Brandt}},
  \bibinfo {author} {\bibfnamefont {T.}~\bibnamefont
  {Bry\ifmmode~\acute{s}\else \'{s}\fi{}}}, \bibinfo {author} {\bibfnamefont
  {M.}~\bibnamefont {Daum}}, \bibinfo {author} {\bibfnamefont {P.}~\bibnamefont
  {Fierlinger}}, \bibinfo {author} {\bibfnamefont {P.}~\bibnamefont {Hautle}},
  \bibinfo {author} {\bibfnamefont {R.}~\bibnamefont {Henneck}}, \bibinfo
  {author} {\bibfnamefont {S.}~\bibnamefont {Heule}}, \bibinfo {author}
  {\bibfnamefont {K.}~\bibnamefont {Kirch}}, \bibinfo {author} {\bibfnamefont
  {J.}~\bibnamefont {Kohlbrecher}}, \bibinfo {author} {\bibfnamefont
  {G.}~\bibnamefont {K\"uhne}}, \bibinfo {author} {\bibfnamefont {J.~A.}\
  \bibnamefont {Konter}}, \bibinfo {author} {\bibfnamefont {A.}~\bibnamefont
  {Pichlmaier}}, \bibinfo {author} {\bibfnamefont {A.}~\bibnamefont {Wokaun}},
  \bibinfo {author} {\bibfnamefont {K.}~\bibnamefont {Bodek}}, \bibinfo
  {author} {\bibfnamefont {M.}~\bibnamefont {Kasprzak}}, \bibinfo {author}
  {\bibfnamefont {M.}~\bibnamefont {Ku\ifmmode~\acute{z}\else \'{z}\fi{}niak}},
  \bibinfo {author} {\bibfnamefont {P.}~\bibnamefont {Geltenbort}},\ and\
  \bibinfo {author} {\bibfnamefont {J.}~\bibnamefont {Zmeskal}},\ }\bibfield
  {title} {\bibinfo {title} {Measured total cross sections of slow neutrons
  scattered by solid deuterium and implications for ultracold neutron
  sources},\ }\href {https://doi.org/10.1103/PhysRevLett.95.182502} {\bibfield
  {journal} {\bibinfo  {journal} {Phys. Rev. Lett.}\ }\textbf {\bibinfo
  {volume} {95}},\ \bibinfo {pages} {182502} (\bibinfo {year}
  {2005}{\natexlab{a}})}\BibitemShut {NoStop}%
\bibitem [{\citenamefont {D\"{o}ge}(2019)}]{doege:2019-phd}%
  \BibitemOpen
  \bibfield  {author} {\bibinfo {author} {\bibfnamefont {S.}~\bibnamefont
  {D\"{o}ge}},\ }\emph {\bibinfo {title} {{S}cattering of {U}ltracold
  {N}eutrons in {C}ondensed {D}euterium and on {M}aterial {S}urfaces}},\ \href
  {https://doi.org/10.14459/2019md1464401} {\bibinfo {type} {{Ph.D.} thesis}},\
  \bibinfo  {school} {Technische {U}niversit\"{a}t M\"{u}nchen, Munich,
  Germany} (\bibinfo {year} {2019}),\ \bibinfo {note}
  {http://doi.org/10.14459/2019md1464401}\BibitemShut {NoStop}%
\bibitem [{\citenamefont {Hamermesh}\ and\ \citenamefont
  {Schwinger}(1946)}]{hamermesh:1946}%
  \BibitemOpen
  \bibfield  {author} {\bibinfo {author} {\bibfnamefont {M.}~\bibnamefont
  {Hamermesh}}\ and\ \bibinfo {author} {\bibfnamefont {J.}~\bibnamefont
  {Schwinger}},\ }\bibfield  {title} {\bibinfo {title} {The {S}cattering of
  {S}low {N}eutrons by {O}rtho- and {P}aradeuterium},\ }\href@noop {}
  {\bibfield  {journal} {\bibinfo  {journal} {Phys. Rev.}\ }\textbf {\bibinfo
  {volume} {69}},\ \bibinfo {pages} {145} (\bibinfo {year} {1946})}\BibitemShut
  {NoStop}%
\bibitem [{\citenamefont {Young}\ and\ \citenamefont
  {Koppel}(1964)}]{young-koppel:1964}%
  \BibitemOpen
  \bibfield  {author} {\bibinfo {author} {\bibfnamefont {J.~A.}\ \bibnamefont
  {Young}}\ and\ \bibinfo {author} {\bibfnamefont {J.~U.}\ \bibnamefont
  {Koppel}},\ }\bibfield  {title} {\bibinfo {title} {{S}low {N}eutron
  {S}cattering by {M}olecular {H}ydrogen and {D}euterium},\ }\href
  {https://doi.org/10.1103/PhysRev.135.A603} {\bibfield  {journal} {\bibinfo
  {journal} {Phys. Rev.}\ }\textbf {\bibinfo {volume} {135}},\ \bibinfo {pages}
  {A603} (\bibinfo {year} {1964})}\BibitemShut {NoStop}%
\bibitem [{\citenamefont {Hill}\ and\ \citenamefont
  {Lounasmaa}(1959)}]{hill:1959}%
  \BibitemOpen
  \bibfield  {author} {\bibinfo {author} {\bibfnamefont {R.~W.}\ \bibnamefont
  {Hill}}\ and\ \bibinfo {author} {\bibfnamefont {O.~V.}\ \bibnamefont
  {Lounasmaa}},\ }\bibfield  {title} {\bibinfo {title} {The lattice specific
  heats of solid hydrogen and deuterium},\ }\href
  {https://doi.org/10.1080/14786435908238235} {\bibfield  {journal} {\bibinfo
  {journal} {Philos. Mag.}\ }\textbf {\bibinfo {volume} {4}},\ \bibinfo {pages}
  {785} (\bibinfo {year} {1959})}\BibitemShut {NoStop}%
\bibitem [{\citenamefont {Nielsen}(1973)}]{nielsen:1973}%
  \BibitemOpen
  \bibfield  {author} {\bibinfo {author} {\bibfnamefont {M.}~\bibnamefont
  {Nielsen}},\ }\bibfield  {title} {\bibinfo {title} {Phonons in solid hydrogen
  and deuterium studied by inelastic coherent neutron scattering},\ }\href
  {https://doi.org/10.1103/PhysRevB.7.1626} {\bibfield  {journal} {\bibinfo
  {journal} {Phys. Rev. B}\ }\textbf {\bibinfo {volume} {7}},\ \bibinfo {pages}
  {1626} (\bibinfo {year} {1973})}\BibitemShut {NoStop}%
\bibitem [{\citenamefont {Ignatovich}(1990)}]{ignatovich:1990}%
  \BibitemOpen
  \bibfield  {author} {\bibinfo {author} {\bibfnamefont {V.~K.}\ \bibnamefont
  {Ignatovich}},\ }\href@noop {} {\emph {\bibinfo {title} {The Physics of
  Ultracold Neutrons}}}\ (\bibinfo  {publisher} {Clarendon Press, Oxford},\
  \bibinfo {year} {1990})\BibitemShut {NoStop}%
\bibitem [{\citenamefont {Golub}\ \emph {et~al.}(1991)\citenamefont {Golub},
  \citenamefont {Richardson},\ and\ \citenamefont {Lamoreaux}}]{golub:1991}%
  \BibitemOpen
  \bibfield  {author} {\bibinfo {author} {\bibfnamefont {R.}~\bibnamefont
  {Golub}}, \bibinfo {author} {\bibfnamefont {D.~J.}\ \bibnamefont
  {Richardson}},\ and\ \bibinfo {author} {\bibfnamefont {S.~K.}\ \bibnamefont
  {Lamoreaux}},\ }\href@noop {} {\emph {\bibinfo {title} {Ultra-Cold
  Neutrons}}}\ (\bibinfo  {publisher} {Adam Hilger, Bristol},\ \bibinfo {year}
  {1991})\BibitemShut {NoStop}%
\bibitem [{\citenamefont {Placzek}\ \emph {et~al.}(1951)\citenamefont
  {Placzek}, \citenamefont {Nijboer},\ and\ \citenamefont
  {Van~Hove}}]{placzek:1951}%
  \BibitemOpen
  \bibfield  {author} {\bibinfo {author} {\bibfnamefont {G.}~\bibnamefont
  {Placzek}}, \bibinfo {author} {\bibfnamefont {B.~R.~A.}\ \bibnamefont
  {Nijboer}},\ and\ \bibinfo {author} {\bibfnamefont {L.}~\bibnamefont
  {Van~Hove}},\ }\bibfield  {title} {\bibinfo {title} {Effect of short
  wavelength interference on neuteron scattering by dense systems of heavy
  nuclei},\ }\href {https://doi.org/10.1103/PhysRev.82.392} {\bibfield
  {journal} {\bibinfo  {journal} {Phys. Rev.}\ }\textbf {\bibinfo {volume}
  {82}},\ \bibinfo {pages} {392} (\bibinfo {year} {1951})}\BibitemShut
  {NoStop}%
\bibitem [{\citenamefont {Gurevich}\ and\ \citenamefont
  {Tarasov}(1968)}]{tarasov-gurevich:1968}%
  \BibitemOpen
  \bibfield  {author} {\bibinfo {author} {\bibfnamefont {I.~I.}\ \bibnamefont
  {Gurevich}}\ and\ \bibinfo {author} {\bibfnamefont {L.~V.}\ \bibnamefont
  {Tarasov}},\ }\href@noop {} {\emph {\bibinfo {title} {Low-Energy Neutron
  Physics}}}\ (\bibinfo  {publisher} {North-Holland Publishing Company},\
  \bibinfo {year} {1968})\BibitemShut {NoStop}%
\bibitem [{\citenamefont {Lovesey}(1986)}]{lovesey:1986}%
  \BibitemOpen
  \bibfield  {author} {\bibinfo {author} {\bibfnamefont {S.~W.}\ \bibnamefont
  {Lovesey}},\ }\href@noop {} {\emph {\bibinfo {title} {Theory of Neutron
  Scattering from Condensed Matter}}},\ Vol.~\bibinfo {volume} {1}\ (\bibinfo
  {publisher} {Oxford Science Publications},\ \bibinfo {year}
  {1986})\BibitemShut {NoStop}%
\bibitem [{\citenamefont {Squires}(1978)}]{squires:1978}%
  \BibitemOpen
  \bibfield  {author} {\bibinfo {author} {\bibfnamefont {G.~L.}\ \bibnamefont
  {Squires}},\ }\href@noop {} {\emph {\bibinfo {title} {Introduction to the
  Theory of Thermal Neutron Scattering}}}\ (\bibinfo  {publisher} {Dover
  Publications, Mineola, New York},\ \bibinfo {year} {1978})\BibitemShut
  {NoStop}%
\bibitem [{\citenamefont {Stepanov}(1975)}]{stepanov:1975}%
  \BibitemOpen
  \bibfield  {author} {\bibinfo {author} {\bibfnamefont {A.~V.}\ \bibnamefont
  {Stepanov}},\ }\bibfield  {title} {\bibinfo {title} {On the coefficient of
  absorption of ultracold neutrons in a medium bounded by a rough surface},\
  }\href {https://doi.org/10.1007/BF01037809} {\bibfield  {journal} {\bibinfo
  {journal} {Theoretical and Mathematical Physics}\ }\textbf {\bibinfo {volume}
  {22}},\ \bibinfo {pages} {299} (\bibinfo {year} {1975})}\BibitemShut
  {NoStop}%
\bibitem [{\citenamefont {Abramowitz}\ and\ \citenamefont
  {Stegun}(1965)}]{abramowitz-stegun:1965}%
  \BibitemOpen
  \bibfield  {author} {\bibinfo {author} {\bibfnamefont {M.}~\bibnamefont
  {Abramowitz}}\ and\ \bibinfo {author} {\bibfnamefont {I.~A.}\ \bibnamefont
  {Stegun}},\ }\href@noop {} {\emph {\bibinfo {title} {Handbook of Mathematical
  Functions}}}\ (\bibinfo  {publisher} {Dover Publications, Mineola, New
  York},\ \bibinfo {year} {1965})\BibitemShut {NoStop}%
\bibitem [{\citenamefont {Placzek}\ and\ \citenamefont
  {Van~Hove}(1955)}]{placzek:1955}%
  \BibitemOpen
  \bibfield  {author} {\bibinfo {author} {\bibfnamefont {G.}~\bibnamefont
  {Placzek}}\ and\ \bibinfo {author} {\bibfnamefont {L.}~\bibnamefont
  {Van~Hove}},\ }\bibfield  {title} {\bibinfo {title} {Interference effects in
  the total neutron scattering cross-section of crystals},\ }\href
  {https://doi.org/10.1007/BF02731767} {\bibfield  {journal} {\bibinfo
  {journal} {Il Nuovo Cimento (1955-1965)}\ }\textbf {\bibinfo {volume} {1}},\
  \bibinfo {pages} {233} (\bibinfo {year} {1955})}\BibitemShut {NoStop}%
\bibitem [{\citenamefont {Oskotskii}(1967)}]{oskotskii:1967}%
  \BibitemOpen
  \bibfield  {author} {\bibinfo {author} {\bibfnamefont {V.~S.}\ \bibnamefont
  {Oskotskii}},\ }\bibfield  {title} {\bibinfo {title} {Measurement of phonon
  distribution function in polycrystalline materials using coherent scattering
  of slow neutrons into a solid angle},\ }\href@noop {} {\bibfield  {journal}
  {\bibinfo  {journal} {Sov. Phys. Solid State}\ }\textbf {\bibinfo {volume}
  {9}},\ \bibinfo {pages} {420} (\bibinfo {year} {1967})},\ \bibinfo {note}
  {[Fiz. Tverd. Tela (Leningrad) 9, 550 (1967)]}\BibitemShut {NoStop}%
\bibitem [{\citenamefont {Schober}(2014)}]{schober:2014}%
  \BibitemOpen
  \bibfield  {author} {\bibinfo {author} {\bibfnamefont {H.}~\bibnamefont
  {Schober}},\ }\bibfield  {title} {\bibinfo {title} {An introduction to the
  theory of nuclear neutron scattering in condensed matter},\ }\href
  {https://doi.org/10.3233/JNR-140016} {\bibfield  {journal} {\bibinfo
  {journal} {Journal of Neutron Research}\ }\textbf {\bibinfo {volume} {17}},\
  \bibinfo {pages} {109} (\bibinfo {year} {2014})}\BibitemShut {NoStop}%
\bibitem [{\citenamefont {D\"oge}\ \emph {et~al.}(2020)\citenamefont {D\"oge},
  \citenamefont {Hingerl}, \citenamefont {Lychagin},\ and\ \citenamefont
  {Morkel}}]{doege:2020-foils}%
  \BibitemOpen
  \bibfield  {author} {\bibinfo {author} {\bibfnamefont {S.}~\bibnamefont
  {D\"oge}}, \bibinfo {author} {\bibfnamefont {J.}~\bibnamefont {Hingerl}},
  \bibinfo {author} {\bibfnamefont {E.~V.}\ \bibnamefont {Lychagin}},\ and\
  \bibinfo {author} {\bibfnamefont {C.}~\bibnamefont {Morkel}},\ }\bibfield
  {title} {\bibinfo {title} {Scattering of ultracold neutrons from rough
  surfaces of metal foils},\ }\href
  {https://doi.org/10.1103/PhysRevC.102.064607} {\bibfield  {journal} {\bibinfo
   {journal} {Phys. Rev. C}\ }\textbf {\bibinfo {volume} {102}},\ \bibinfo
  {pages} {064607} (\bibinfo {year} {2020})}\BibitemShut {NoStop}%
\bibitem [{\citenamefont {Binder}(1970)}]{binder:1970}%
  \BibitemOpen
  \bibfield  {author} {\bibinfo {author} {\bibfnamefont {K.}~\bibnamefont
  {Binder}},\ }\bibfield  {title} {\bibinfo {title} {Total coherent cross
  sections for the scattering of neutrons from crystals},\ }\href
  {https://doi.org/10.1002/pssb.19700410233} {\bibfield  {journal} {\bibinfo
  {journal} {Physica Status Solidi B}\ }\textbf {\bibinfo {volume} {41}},\
  \bibinfo {pages} {767} (\bibinfo {year} {1970})}\BibitemShut {NoStop}%
\bibitem [{\citenamefont {Mucker}\ \emph {et~al.}(1968)\citenamefont {Mucker},
  \citenamefont {Harris}, \citenamefont {White},\ and\ \citenamefont
  {Erickson}}]{mucker:1968}%
  \BibitemOpen
  \bibfield  {author} {\bibinfo {author} {\bibfnamefont {K.~F.}\ \bibnamefont
  {Mucker}}, \bibinfo {author} {\bibfnamefont {P.~M.}\ \bibnamefont {Harris}},
  \bibinfo {author} {\bibfnamefont {D.}~\bibnamefont {White}},\ and\ \bibinfo
  {author} {\bibfnamefont {R.~A.}\ \bibnamefont {Erickson}},\ }\bibfield
  {title} {\bibinfo {title} {Structures of solid deuterium above and below the
  $\lambda$ transition as determined by neutron diffraction},\ }\href
  {https://doi.org/10.1063/1.1670327} {\bibfield  {journal} {\bibinfo
  {journal} {The Journal of Chemical Physics}\ }\textbf {\bibinfo {volume}
  {49}},\ \bibinfo {pages} {1922} (\bibinfo {year} {1968})}\BibitemShut
  {NoStop}%
\bibitem [{\citenamefont {Mucker}\ \emph {et~al.}(1966)\citenamefont {Mucker},
  \citenamefont {Talhouk}, \citenamefont {Harris},\ and\ \citenamefont
  {White}}]{mucker:1966}%
  \BibitemOpen
  \bibfield  {author} {\bibinfo {author} {\bibfnamefont {K.~F.}\ \bibnamefont
  {Mucker}}, \bibinfo {author} {\bibfnamefont {S.}~\bibnamefont {Talhouk}},
  \bibinfo {author} {\bibfnamefont {P.~M.}\ \bibnamefont {Harris}},\ and\
  \bibinfo {author} {\bibfnamefont {D.}~\bibnamefont {White}},\ }\bibfield
  {title} {\bibinfo {title} {Crystal structure of para-enriched solid deuterium
  below the $\lambda$ transition},\ }\href@noop {} {\bibfield  {journal}
  {\bibinfo  {journal} {Phys. Rev. Lett.}\ }\textbf {\bibinfo {volume} {16}},\
  \bibinfo {pages} {799} (\bibinfo {year} {1966})}\BibitemShut {NoStop}%
\bibitem [{\citenamefont {Liu}\ \emph {et~al.}(2010)\citenamefont {Liu},
  \citenamefont {Young}, \citenamefont {Lavelle},\ and\ \citenamefont
  {Salvat}}]{liu:2010}%
  \BibitemOpen
  \bibfield  {author} {\bibinfo {author} {\bibfnamefont {C.-Y.}\ \bibnamefont
  {Liu}}, \bibinfo {author} {\bibfnamefont {A.~R.}\ \bibnamefont {Young}},
  \bibinfo {author} {\bibfnamefont {C.~M.}\ \bibnamefont {Lavelle}},\ and\
  \bibinfo {author} {\bibfnamefont {D.}~\bibnamefont {Salvat}},\ }\bibfield
  {title} {\bibinfo {title} {Coherent neutron scattering in polycrystalline
  deuterium and its implications for ultracold neutron production},\ }\href
  {https://arxiv.org/abs/1005.1016v1} {\bibfield  {journal} {\bibinfo
  {journal} {arXiv preprint}\ } (\bibinfo {year} {2010})},\ \bibinfo {note}
  {arXiv:1005.1016 [nucl-th]}\BibitemShut {NoStop}%
\bibitem [{\citenamefont {Atchison}\ \emph
  {et~al.}(2005{\natexlab{b}})\citenamefont {Atchison}, \citenamefont {van~den
  Brandt}, \citenamefont {Bry\ifmmode~\acute{s}\else \'{s}\fi{}}, \citenamefont
  {Daum}, \citenamefont {Fierlinger}, \citenamefont {Hautle}, \citenamefont
  {Henneck}, \citenamefont {Heule}, \citenamefont {Kasprzak}, \citenamefont
  {Kirch}, \citenamefont {Konter}, \citenamefont {Michels}, \citenamefont
  {Pichlmaier}, \citenamefont {Wohlmuther}, \citenamefont {Wokaun},
  \citenamefont {Bodek}, \citenamefont {Szerer}, \citenamefont {Geltenbort},
  \citenamefont {Zmeskal},\ and\ \citenamefont
  {Pokotilovskiy}}]{atchison:2005-ucnprod}%
  \BibitemOpen
  \bibfield  {author} {\bibinfo {author} {\bibfnamefont {F.}~\bibnamefont
  {Atchison}}, \bibinfo {author} {\bibfnamefont {B.}~\bibnamefont {van~den
  Brandt}}, \bibinfo {author} {\bibfnamefont {T.}~\bibnamefont
  {Bry\ifmmode~\acute{s}\else \'{s}\fi{}}}, \bibinfo {author} {\bibfnamefont
  {M.}~\bibnamefont {Daum}}, \bibinfo {author} {\bibfnamefont {P.}~\bibnamefont
  {Fierlinger}}, \bibinfo {author} {\bibfnamefont {P.}~\bibnamefont {Hautle}},
  \bibinfo {author} {\bibfnamefont {R.}~\bibnamefont {Henneck}}, \bibinfo
  {author} {\bibfnamefont {S.}~\bibnamefont {Heule}}, \bibinfo {author}
  {\bibfnamefont {M.}~\bibnamefont {Kasprzak}}, \bibinfo {author}
  {\bibfnamefont {K.}~\bibnamefont {Kirch}}, \bibinfo {author} {\bibfnamefont
  {J.~A.}\ \bibnamefont {Konter}}, \bibinfo {author} {\bibfnamefont
  {A.}~\bibnamefont {Michels}}, \bibinfo {author} {\bibfnamefont
  {A.}~\bibnamefont {Pichlmaier}}, \bibinfo {author} {\bibfnamefont
  {M.}~\bibnamefont {Wohlmuther}}, \bibinfo {author} {\bibfnamefont
  {A.}~\bibnamefont {Wokaun}}, \bibinfo {author} {\bibfnamefont
  {K.}~\bibnamefont {Bodek}}, \bibinfo {author} {\bibfnamefont
  {U.}~\bibnamefont {Szerer}}, \bibinfo {author} {\bibfnamefont
  {P.}~\bibnamefont {Geltenbort}}, \bibinfo {author} {\bibfnamefont
  {J.}~\bibnamefont {Zmeskal}},\ and\ \bibinfo {author} {\bibfnamefont
  {Y.}~\bibnamefont {Pokotilovskiy}},\ }\bibfield  {title} {\bibinfo {title}
  {Production of ultracold neutrons from a cold neutron beam on a
  $^{2}\text{H}_{2}$ target},\ }\href
  {https://doi.org/10.1103/PhysRevC.71.054601} {\bibfield  {journal} {\bibinfo
  {journal} {Phys. Rev. C}\ }\textbf {\bibinfo {volume} {71}},\ \bibinfo
  {pages} {054601} (\bibinfo {year} {2005}{\natexlab{b}})}\BibitemShut
  {NoStop}%
\end{thebibliography}%

\end{document}